\documentclass{aa}
\usepackage{psfig,latexsym,longtable}

\def\sun{\hbox{$\odot$~}}
\def\deg{\hbox{$^\circ$}}
\def\lefevre{Le\thinspace F\`evre~}

\begin{document}

\thesaurus{03(11.03.1; 11.09.3; 12.03.3; 13.25.2)}

\title{{\it Beppo}SAX observations of two high redshift clusters of galaxies: RXJ\thinspace 0152.7$-$1357 and MS\thinspace 2053.7$-$0449}

\author{R. Della Ceca \inst{1}, R. Scaramella \inst{2}, I.M. Gioia \inst{3,} \inst{4}, 
P. Rosati \inst{5}, F. Fiore \inst {2,} \inst{6} and G. Squires \inst{7,} \inst{8}}

\institute{
Osservatorio Astronomico di Brera, via Brera 28, I-20121 Milano, Italy.
\and
Osservatorio Astronomico di Roma, via Frascati 33, I-00040 Monteporzio Catone, Italy.
\and
Institute for Astronomy, 2680 Woodlawn Drive, Honolulu, HI 96822, USA.
\and 
Istituto di Radioastronomia del CNR, via Gobetti 101, I-40129 Bologna, Italy.
\and
European Southern Observatory, D-85748, Garching bei Munchen, Germany.
\and
BeppoSAX Science Data Center, via Corcolle 19, I-00131 Roma, Italy.
\and
Center for Particle Astrophysics, UC Berkeley, 301 Le Conte Hall, Berkeley, CA, USA.
\and
Caltech Astronomy M/S 105-24, 1200 E. California Blvd., Pasadena, CA, USA.
}

\date{Received: August 4, 1999; Accepted: October 25, 1999}

\authorrunning{Della Ceca et al.,}
\titlerunning{{\it Beppo}SAX observations of two high $z$ clusters of galaxies.}

\maketitle

\begin{abstract}

We present X-ray observations of two high redshift clusters of galaxies
carried out with the {\it Beppo}SAX satellite. One cluster,
RXJ\thinspace 0152.7$-$1357 at $z \simeq 0.83$, was selected from the ROSAT
Deep Cluster Survey sample, as one of the most X-ray luminous systems
known at $z>0.5$. 
The optical and ROSAT-PSPC data show a complex
morphology with at least two cores. Our SAX observations yield a gas
temperature $kT = 6.46^{+1.74}_{-1.19}$ keV and a metallicity  $A
= 0.53^{+0.29}_{-0.24}$, with a prominent iron K$_{\alpha}$ line.  The
second cluster, MS\thinspace 2053.7$-$0449 at $z \simeq 0.58$, was selected
from the EMSS sample. 
Given the poor statistics no constraints on the metallicity can be
derived from the present data.
Large uncertainties are associated to the gas temperature 
($kT = 6.7^{+6.8}_{-2.3}$ keV), which has been obtained after 
fixing the abundance to 0.3 solar.
Combining these results with those obtained for similarly high
redshift ($z \geq 0.5$) clusters of galaxies with 
broad band X-ray
spectra, we discuss the high redshift $L_{bol} - T$ relationship. 
The data can easily be accommodated with a non-evolving (or mildly
evolving) $L_{bol} - T$ relation, a result which, when combined with
the little observed evolution in the bulk of the X-ray cluster
population, gives support to low $\Omega$ cosmological models.

\end{abstract}

\section{Introduction}

High redshift clusters of galaxies have long served as valuable tools to test
theories of cosmic structure formation (e.g. Press \& Schechter 1974,
Peebles et al. 1989, Eke et al.\ 1996, to mention just a few).
Measurements of the temperature of the intra-cluster medium (ICM) in
distant clusters are particularly important since cluster temperatures  
are closely related to their mass, as long as these systems are in virial
equilibrium. 
Physical cluster parameters of the highest redshift
clusters, measurable by means of deep X-ray observations 
(e.g. the Luminosity-Temperature relation), set also
strong constraints on the thermodynamical evolution
of their gaseous component.
In addition, the measurement of the metal abundance from X-ray spectra in
remote bound systems can constrain the mode and epoch of the enrichment
of the intracluster gas.

Despite their importance, ICM properties at high redshift ($z>0.5$) are
still largely unexplored. This is due to the limited effective area and
spatial resolution of the past X-ray missions, as well as to the
difficulty of finding X-ray luminous clusters at $z>0.5$.
Indeed  the  {\it Extended Medium Sensitivity Survey} (EMSS; Gioia et
al., 1990) carried out with the Einstein Observatory has detected
clusters at redshift up to $\sim 0.8$ (e.g. MS1137+6625 at 0.78 and
MS1054-0321 at z=0.83, Gioia and Luppino, 1994; Donahue et al., 1998) but
the statistics of high-z clusters in the EMSS is still low with only 6
clusters (out of a total of about 100) at $z > 0.5$.
Over the last few years, however, several X-ray cluster surveys based
on ROSAT-PSPC data (the RDCS: Rosati et al., 1995, 1998; the  WARPS:
Scharf et al.,  1997; the SHARC: Burke et al., 1997; the 160$\deg$
survey of Vikhlinin et al.,  1998; the NEP: Mullis et al., 1998) have
changed the observational scenario by collecting sizable samples of
distant ($z>0.5$) systems, thus extending the pioneering work of the
EMSS.  The ROSAT Deep Cluster Survey (RDCS, Rosati et al. 1995, 1998)
has pushed these studies to the faintest fluxes and highest redshifts,
with 33 clusters spectroscopically identified to date at $z>0.5$, out
to $z=1.26$ (Rosati et al. 1999).

As an alternative method, several studies have tried to locate distant
clusters around high redshift radio galaxies (e.g. 3C~324 at $z=1.2$, 
Smail \& Dickinson, 1995; 3C~184 at $z=0.99$, Deltorn et al. 1997). 
ROSAT observations of distant radio galaxies (Crawford \&
Fabian 1996; Dickinson 1997) have provided some evidence that the
observed X-ray emission might originate from hot intra-cluster gas.  It
remains difficult, however, to reliably measure ICM physical parameters
in these objects, since one needs to discriminate an absorbed power-law
component, due to the radio galaxy's AGN, from the thermal component of
the putative diffuse ICM.

Although the number of {\it bona-fide} high-$z$ clusters has
significantly increased, only long ASCA observations, 
with a broad energy response, have allowed
a few sketchy studies of their ICM properties to date.
A recent compilation of cluster temperatures 
(Wu, Xue and Fang, 1999) reports a total of 168 clusters with good 
broad band X-ray spectroscopy, but only 5 (3) of these have been studied at 
$z>0.5$ ($z>0.8$).

In this paper we present {\it Beppo}SAX observations of two high
redshift clusters:  MS\thinspace 2053.7$-$0449 from the EMSS sample at
$z=0.583$ and RXJ\thinspace 0152.7$-$1357 from the RDCS sample at
$z=0.831$.  
The {\it Beppo}SAX MECS instrument (see section 2) has a slightly
smaller (larger) effective area than the ASCA GIS for $E < 7$ keV ($E >
7$ keV) and it is characterized by a significantly sharper PSF if
compared with the ASCA GIS or ASCA SIS instruments, especially at high
energy.  This latter property implies a reduced background, making {\it
Beppo}SAX MECS the best instrument to date for distant clusters
investigations.
RXJ\thinspace 0152.7$-$1357 is the highest redshift cluster with good
quality broad band X-ray spectroscopy to date and among the most X-ray
luminous known at $z>0.6$ together with the well known MS1054.4$-$0321
(Gioia and Luppino, 1994; Donahue et al., 1998). 
Our data provide a rather accurate ($\sim\! 20\%$) temperature
measurement for RXJ\thinspace 0152.7$-$1357, which allows an
interesting comparison with MS1054.4$-$0321 with very similar
redshift and luminosity.
In Table 1 the main properties of the two 
clusters, in their own discovery bands, are summarized.

\begin{table*}
\begin{center}
\caption{Observed Clusters}
\begin{tabular}{lcccccc}
\hline
\hline
Name       & RA       & DEC & z & Energy Band & $f_{x}^{a}$ & $L_{x}^{a}$  \\
& (J2000)  & (J2000) &     & keV & erg cm$^{-2}$ s$^{-1}$  & $\rm{h}_{50}^{-2}$ erg s$^{-1}$  \\
\hline
\hline
RXJ\thinspace 0152.7$-$1357  & 01 52 41  & -13 57 45  & 0.831  & 0.5 $-$ 2.0   & $2.2\pm 0.2 \times 10^{-13}$   & $6.8\pm 0.6 \times 10^{44}$ \\
MS\thinspace  2053.7$-$0449  & 20 56 22  & -04 37 52  & 0.583  & 0.3 $-$ 3.5   & $4.0\pm 0.9 \times 10^{-13}$   & $5.8\pm 1.3 \times 10^{44}$ \\
\hline
\hline
\end{tabular}
\end{center}
$\ ^{a} $ We note that in the case of RXJ\thinspace 0152.7$-$1357 we
have reported the  ROSAT PSPC flux$_{(0.5 - 2.0\  keV)}$ (luminosity)
measured over a circle of 3 arcmin radius, corresponding to  $1.5
h_{50}^{-1}$ Mpc at $z=0.83$, while for MS\thinspace  2053.7$-$0449 we
have reported the total EMSS ``corrected" flux$_{(0.3 - 3.5\ keV)}$
(luminosity) i.e.  the flux (luminosity) corrected for the effect of
the finite EMSS detection cell.
\end{table*}

This paper is organized as follows.
In section 2 we present the {\it Beppo}SAX observations. The spectral analysis 
for RXJ\thinspace 0152.7$-$1357 is presented in section 3 and for
MS\thinspace 2053.7$-$0449 in section 4. Section 5 contains a discussion 
of the results in the context of the high redshift $L_{bol} - T$ relationship.
Finally, summary and conclusions are presented in section 6. We adopt 
$H_{0} = 50$ km s$^{-1}$ Mpc$^{-1}$ and $q_{0} = 0.5$ throughout, unless 
otherwise noted.

\section {{\it Beppo}SAX Observations and Data Preparation}
           
\begin{table*}
\begin{center}
\caption{{\it Beppo}SAX Data Observation Journal}
\begin{tabular}{lcccccc}
\hline
\hline
Name & Date & Seq. Numb & Exp. LECS & Rate LECS         &  Exp. MECS & Rate MECS         \\
     &      &           &  ksec    & (10$^{-3}$ cts/s) &  ksec      & (10$^{-3}$ cts/s)  \\
     &      &           &          & (0.12 $-$ 4.0 keV) &            & (1.65 $-$ 10 keV)    \\ 
\hline  
\hline
RXJ\thinspace 0152.7$-$1357 & 1998 Aug 7-10 & 60605001 & 61.3 & $2.7 \pm 0.3$ & 127.7 & $4.0 \pm 0.3$  \\
MS\thinspace 2053.7$-$0449  & 1997 Oct 21-23 & 60243001 & 32.7 & $2.7  \pm 0.4$ & 81.9 & $2.4 \pm 0.3$  \\
\hline
\hline
\end{tabular}
\end{center}
\end{table*}

The observations were performed with the {\it Beppo}SAX Narrow Field
Instruments, LECS (0.1--10 keV, Parmar et al. 1997), MECS (1.3--10 keV,
Boella et al.  1997), HPGSPC (4--60 keV, Manzo et al. 1997) and PDS
(13--200 keV, Frontera et al. 1997). We report here the analysis of the
imaging instruments data (LECS and MECS). PDS and HPGSPC are collimated
instruments with field of view of about 1.5 degrees (FWHM) and have a
rather large and structured background which makes them not sensitive
enough for very faint sources.

At launch the MECS was composed of three identical units. Unfortunately
on 1997 May 6$^{\mathit{th}}$ the unit MECS1 had 
a technical failure.  
All observations after this date were performed with two
units (MECS2 and MECS3). The LECS is operated during spacecraft dark
time only, therefore LECS exposure times are usually smaller than MECS
exposures by a factor 1.5-3. The MECS energy resolution is about 
8\% at 6 keV, while the LECS energy resolution is about 11\% at 3 keV.

Table 2 gives the journal of observations for the two high redshift 
clusters. For both clusters we have acquired $\approx 30 \%$ fewer 
photons than anticipated (because of the unavailability of one of the 
three MECS in the case of MS$\thinspace 2053.7.7-0449$, and because
of the shorter effective exposure in the case of 
RXJ\thinspace 0152.7$-$1357).

Standard data reduction was performed using the software package
``SAXDAS'' (see http://www.sdc.asi.it/software and Fiore, Guainazzi \&
Grandi 1999). In particular, data are linearized and cleaned from Earth
occultation periods and unwanted periods of high particle background
(satellite passages through the South Atlantic Anomaly). We accumulated
data for Earth elevation angles $>5$ degrees and magnetic cut-off
rigidity $>6$.  Data from the two MECS units have been merged
after gain equalization and single MECS spectra have been accumulated.

Both MECS and LECS source 
counts have been extracted from a circular region of 4 arcmin radius 
to maximize the statistics and the signal-to-noise (S/N) ratio.

LECS and MECS internal backgrounds depend on the position of the 
target in the detector (see
Chiappetti et al. 1998, the {\it Beppo}SAX Cookbook,
http://www.sdc.asi.it/software/cookbook and Parmar et al. 1999).
Accordingly, background spectra were extracted from high Galactic
latitude ``blank" fields  from the same source extraction
region, in detector coordinates. The mean level of the background 
in the ``blank fields" was compared with that of our observations
using source-free regions at various positions in the detectors.
The level of the ``blank fields'' background is consistent
with the ``local'' background in the two cluster observations.

Spectral fits were performed using the XSPEC 9.0 software package and
the September 1997 version of the calibration files 
(Ancillary Response Files  and Redistribution Matrix Files).
In the spectra analysis only the 1.65$-$10 keV band counts for the MECS
(channels 37 $-$ 220) and the 0.12$-$4 keV counts for the LECS
(channels 11 $-$ 400) were used as suggested by the {\it Beppo}SAX
Cookbook (Fiore, Guainazzi \& Grandi, 1999, v. 1.2).
The background subtracted count rates ($\pm 1 \sigma$) in both LECS 
(0.12-4 keV) and MECS (1.65-10 keV) are given in Table 2.
Source counts were binned in order to have a S/N $\ge$ 3 in each energy
bin.  The LECS and MECS data were fitted jointly, and the model
normalization for each data set were allowed to be an independent
parameter, in order to take into account differences in the absolute
calibration of the two instruments.  If not explicitly quoted all the
errors reported in this paper represent the $68\%$ confidence levels
for 1 interesting parameter ($\Delta \chi^{2}$ = 1.0).
%
%

\section {RXJ\thinspace 0152.7$-$1357}

RXJ\thinspace 0152.7$-$1357 was discovered in the 
RDCS  which 
has constructed an X-ray selected, flux-limited sample of clusters of 
galaxies via
a serendipitous search for extended X-ray sources in deep pointed PSPC
observations. A wavelet-based technique was employed to characterize
low-surface brightness sources and select cluster candidates down to the
flux limit  $f_{0.5-2.0 keV}$ of $1\times
10^{-14}$ erg cm$^{-2}$ s$^{-1}$, over $\sim\! 50$ deg$^2$
(Rosati et al., 1995).
Optical
follow-up imaging and spectroscopy have confirmed to date $\sim\! 140$
clusters and groups, which span a large range in redshift [0.05--1.26]
and X-ray luminosity [$1 \times 10^{42} - 8 \times 10^{44} \rm{h}_{50}^{-2}$ 
erg s$^{-1}$], RXJ\thinspace 0152.7$-$1357 being the most
luminous of the sample.

RXJ\thinspace 0152.7$-$1357 was discovered in the ROSAT PSPC field
rp600005n00 pointed at the nearby galaxy NGC 720. 
The cluster
was detected at 13.6$'$ off-axis ($\alpha_{2000}=$
$01^{h}52^{m}44^{s}$ and $\delta_{2000}= -13^{o}57'21"$).  At this
position the wavelet algorithm clearly reveals an extended X-ray source
with an extent $6.5 \sigma$ above the local PSF of the ROSAT PSPC
instrument and with a double core (the multi-scale wavelet analysis
preserves source substructure information) with intensity peaks separated 
by 
$\sim 1.2$ arcmin ($\sim 0.6  h_{50}^{-1}$ Mpc at the cluster redshift of $z=0.83$). 
This same cluster has also  independently been identified in the WARPS survey
(Ebeling et al. 1999) and recently reported also in Romer et al. (1999).

Integrating the ROSAT PSPC flux over a circle of 3 arcmin radius,
corresponding to $1.5 h_{50}^{-1}$ Mpc at $z=0.83$, we obtain $377 \pm 27$
net counts 
\footnote{
The south-west faint source, visible in figure 1 ($\Delta \alpha \sim
1.9^{\prime}$ and $\Delta \delta \sim -1.8^{\prime}$), possibly an
AGN, was masked out in this measurement (this would increase
the flux by 7\%).  The ROSAT hardness ratio map indicates that the
spectra of this faint source is steeper than that of RXJ\thinspace
0152.7$-$1357; therefore its contribution to the {\it Beppo}SAX data at
$E > 1$ kev is below a few percents.  An even smaller contribution is
expected from the two bright stars visible in the south-east section of the
image.
          }
in the 0.5-2.0 keV energy band, with 6.2/arcmin$^2$ background counts. 
With an effective exposure time of 19904 sec, a galactic HI column
density along the line of sight of  $1.55\times 10^{20} \rm{cm}^{-2}$ 
 (Dickey \& Lockman 1990) 
and using a Raymond-Smith spectral model 
 (Raymond \& Smith 1977)
with kT $\sim$ 6 keV,
we obtain an unabsorbed flux of $f_{0.5 - 2.0 keV} =  (2.2 \pm 0.2)
\times 10^{-13}$ erg cm$^{-2}$ s$^{-1}$.
The rest frame X-ray luminosity within an aperture of 
$1.5 h_{50}^{-1}$ Mpc is $L_X= (6.8\pm 0.6) \times 10^{44} \rm{h}_{50}^{-2}$ erg s$^{-1}$. 

Imaging and spectroscopic follow-up observations of  RXJ1052.7$-$1357
were conducted with the EFOSC1 spectrograph at the ESO 3.6m in November
1996.  
With a long slit exposure of 2 hours
it was possible to secure redshifts for 3 galaxies with
$<z> \simeq 0.831$  belonging to the main northern clump.  Additional
multiband imaging and spectroscopy have been carried out at Keck with
the Low Resolution Imaging Spectrograph (LRIS, Oke et al., 1995) 
and are presented and
discussed elsewhere (Squires et al. in preparation), along with a weak lensing
analysis of the field.  We only note here that the new
spectroscopic data confirm the redshift of $z \sim 0.83$ also for the
southern clump.
 
In figure 1, we show the R-band image (obtained with Keck-LRIS; 4000 s
integration time) with the PSPC X-ray contours overlaid. The cluster
galaxies distribution shows two main clumps and correlates very well
with the double-peaked X-ray morphology.

\begin{figure*}
\hskip 20mm \psfig{file=9124.dummy,height=15.0cm}
\caption{R band image of RXJ1052.7$-$1357
with overlaid ROSAT PSPC X-ray contours at 3 (dashed line), 5, 10, 20 and 30 
$\sigma$
above the background. 
The image was obtained with Keck-LRIS with 4000 s integration time 
(Squires et al. in preparation). 
The right and top axes give the physical linear scale at $z = 0.831$.
The center coordinates are $\alpha_{2000}=$ $01^{h}52^{m}41^{s}$; 
$\delta_{2000}= -13^{o}57'45"$; north is up, east to the left.
}
\end{figure*}

The {\it Beppo}SAX - MECS X-ray image of the field in the $2-10$ keV band is 
shown in figure 2. The 
source, centered at $\alpha_{2000}=$ $01^{h}52^{m}41^{s}$ and 
$\delta_{2000}= -13^{o}57'23"$, is consistent with the
celestial 
position of the cluster within the SAX positional errors.
This source is marginally resolved at the spatial resolution of the 
{\it Beppo}SAX - MECS.

\begin{figure}
\vskip 1.5 true cm
\psfig{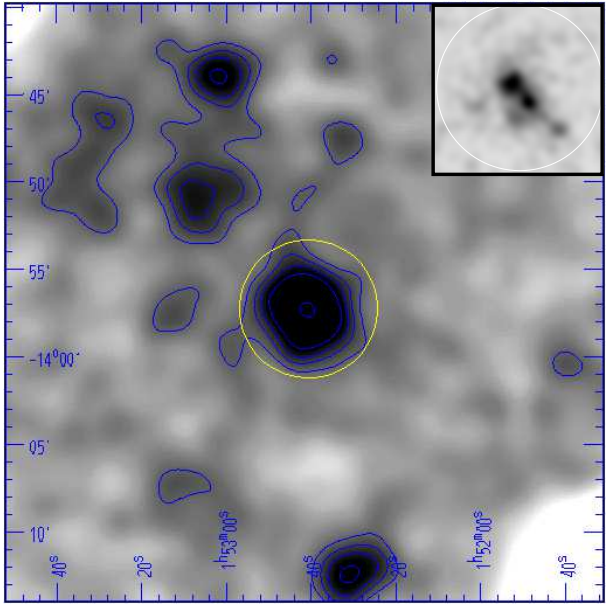}
\vskip -0.7 true cm
\caption{MECS image in the $2-10$ keV band of the RXJ\thinspace 0152.7$-$1357 field
($35' \times 35'$). The raw data have been smoothed with a
gaussian filter with $\sigma = 0.8'$.
Contours are at 3, 5, 7, 10 and 20 sigma above the background.
The white $4\arcmin$ radius circle represents the {\it Beppo}SAX 
counts extraction region for
RXJ\thinspace 0152.7$-$1357. The inset shows the ROSAT-PSPC cutout of 
the same central region in the [0.5-2.0] keV band, smoothed with a gaussian 
with $\sigma \simeq 14\arcsec$. 
The inset clearly shows the double core nature of RXJ\thinspace 0152.7$-$1357.
}
\end{figure}

The total net counts in the source extraction region are $165 \pm 17$
for the LECS and $506 \pm 36$ for the MECS. The net counts represent
about 59\% (39\%) of the total (source + background) counts in the source region for LECS
(MECS).

The source spectrum was fitted using a Raymond-Smith 
spectral model modified by galactic absorption
($N_{H}=1.55\times10^{20}$ cm$^{-2}$) 
along the line of sight at the cluster position.  
The results are reported in Table 3.  The cluster gas
is best fitted by a rest frame temperature  of $kT =
6.46^{+1.74}_{-1.19}$ and metallicity of $A = 0.53^{+0.29}_{-0.24}$
(68\% confidence interval).
The ratio of the model normalizations from LECS and MECS 
is about 0.71, which is consistent with the known differences in the 
absolute calibration of the two instruments.
 
The unfolded spectrum, the folded spectrum and the ratio between the
data and best fit model are displayed in figure 3, while in figure 4 we
report the two-parameter $\chi^{2}$ contours (68.3\%, 90\% and 99\%
confidence levels) for the cluster metallicity (in solar units) and
temperature (in keV).  Both temperature and metal abundance are rather
well constrained, which is partially due to the detection of the iron
$K \alpha$ complex at the redshifted energy of about 3.7 keV.  It is
worth noting that leaving the redshift, along with the metallicity and
temperature, as free parameters we obtain best fit values 
($\chi^{2}_{\nu} / dof = 0.82/20$)
of  $z =
0.80^{+0.07}_{-0.05}$, $kT = 6.16^{+1.93}_{-1.08}$ and metallicity $A =
0.53^{+0.29}_{-0.24}$ (68\% confidence interval).
Finally, we have also tried a fit with the absorbing column density as a 
free parameter. We obtain best fit values 
($\chi^{2}_{\nu} / dof = 0.84/19$)
of 
$kT = 5.98^{+1.68}_{-1.37}$, 
$A = 0.51^{+0.27}_{-0.23}$ and 
$z =0.81^{+0.06}_{-0.06}$;
the best fit absorbing column density converges to zero with a 
$1 \sigma$ upper limit of $2.4 \times 10^{21}$ cm$^{-2}$.
These results confirm the robustness of our spectral fit.

\begin{figure}
\hskip -1truecm
\psfig{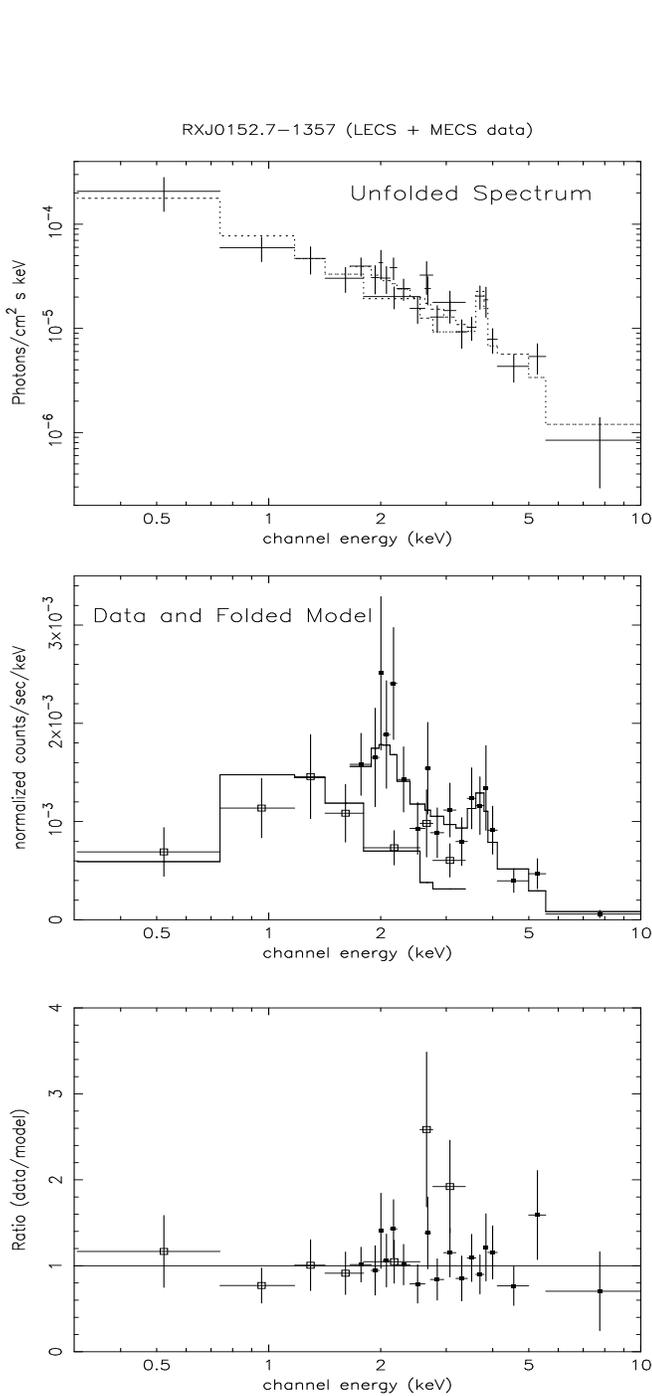}
\vskip 1.7truecm
\caption{RXJ\thinspace 0152.7$-$1357: Unfolded spectrum, folded
spectrum and the ratio between the data and best fit model.  In the
last two panels we have indicated the LECS (MECS) data as open
(filled) squares.  A prominent iron K$_{\alpha}$ complex at the
redshifted energy of $\sim 3.7$ keV is visible.}
\end{figure}

\begin{figure}
\psfig{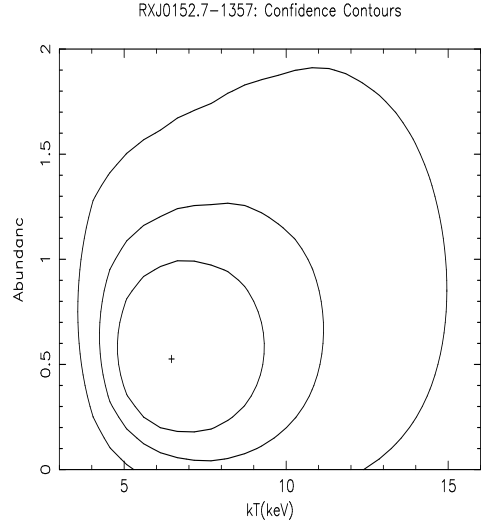}
\caption{RXJ\thinspace 0152.7$-$1357: Two dimensional $\chi^{2}$ 
contours at 68.3\%, 90\% and 99\% confidence levels 
($\Delta \chi^{2}$ = 2.3, 4.61 and 9.21) for the cluster abundance 
(solar units) and temperature (in keV).}
\end{figure}

Using the best fit model parameters reported in Table 3 and assuming a
mean value between the LECS and MECS normalizations, we derive an
unabsorbed ($0.5 - 2.0$) keV flux of $f_{0.5 - 2 keV} = (1.9 \pm 0.4)
\times 10^{-13}$ erg cm$^{-2}$ s$^{-1}$.  The measured flux is in good
agreement with that derived from the ROSAT PSPC data (see Table 1).

The derived unabsorbed ($2 - 10$) keV flux is 
$f_{2 - 10 keV} =  (2.3 \pm 0.5) \times 10^{-13}$ erg cm$^{-2}$ s$^{-1}$, 
corresponding to a luminosity in the cluster rest frame of 
$L_{2 - 10 keV} =  (1.1 \pm 0.2) \times 10^{45} \rm{h}_{50}^{-2}$ erg s$^{-1}$, and 
to a bolometric luminosity of $L_{bol} =  (2.2 \pm 0.5) \times 10^{45} \rm{h}_{50}^{-2}$
erg s$^{-1}$.

Given the uncertain dynamical state of RXJ0152.7-1357 and the difficulty in measuring
its gas density profile, it is difficult to provide an accurate estimate of its mass
\footnote{ 
Assuming the gas in isothermal and hydrostatic equilibrium
the total mass within a radius $r (r >> r_c)$
would be :
$$ M(<r) \simeq 10^{15} \left({T\over 6.5\,\rm{keV}}\right)
 \left({r\over \rm{Mpc}}\right)
 \left({\beta\over 0.7}\right) h_{50}^{-1} M_{\odot} $$
}
.


\section {MS\thinspace 2053.7$-$0449}

MS\thinspace 2053.7$-$0449 is part of the EMSS sample of high-$z$ clusters of 
galaxies ($z>0.5$) serendipitously discovered in the fields of the 
Einstein IPC. The X-ray source is located at $\alpha_{2000}=$ 
$20^{h}56^{m}22^{s}$ and $\delta_{2000}= -04^{o}37'52"$ and the unabsorbed
IPC flux, corrected for the effect of the finite EMSS detection cell 
is $f_{0.3 - 3.5 keV} = (4.0  \pm 0.9) \times 10^{-13}$ erg cm$^{-2}$ s$^{-1}$
(Henry et al. 1992; Gioia \& Luppino 1994). The redshift of $z=0.583$ is 
based  on 5 concordant galaxy redshifts (Fabricant, private 
communication). 

Given its high luminosity, 
$L_{0.3 - 3.5 keV} =  (5.78 \pm 1.3) \times 10^{44} \rm{h}_{50}^{-2}$ erg s$^{-1}$, 
MS\thinspace 2053.7$-$0449 was part of the CCD imaging survey for 
gravitational  lensing carried out with the telescopes at Mauna Kea by  
Luppino et al. (1999).  Luppino and Gioia (1992) first reported the 
presence of a large arc (arc-length $\sim11''$) fragmented into two 
distinct clumps at a radius of $\sim16''$ from the optically dominant 
cluster galaxy. 

A weak lensing study (Clowe 1998) shows that  
MS\thinspace 2053.7$-$0449 is not among the most massive
z$\sim$ 0.55 clusters of the EMSS. Clowe et al. (1999) report
a mass value, from the weak lensing, of 
(2.3 $\pm 1.1)\times10^{14}$h$_{100}^{-1} M_{\sun}$ within 0.5 
h$^{-1}_{100}$ Mpc. The mass profile is well fit by a ``universal'' CDM
profile (Navarro, Frenk \& White, 1996) with parameters r$_{200}$=590 h$^{-1}_{100}$
kpc and c$=$2  assuming a background galaxy redshift 
z$_{bg}$=1.5. MS\thinspace 2053.7$-$0449 is also well fit by an 
isothermal sphere model with a velocity dispersion of $\sigma=730$
km s$^{-1}$ for z$_{bg}$=1.5, indicating that the cluster is close
to virialization.
Kelson et al. (1997) find that the fundamental plane relation of galaxies 
in   MS\thinspace 2053.7$-$0449 is very similar to that of Coma, suggesting 
that the structure of the early-type galaxies has changed little since 
$z=0.58$.

The MECS $2-10$ keV X-ray image of the field is shown in figure 5; the
peak of the X-ray emission from the cluster is centered at
$\alpha_{2000}=$ $20^{h}56^{m}21^{s}$ and $\delta_{2000}=
-04^{o}38'53"$, which is consistent with the celestial position of the
cluster within the SAX positional errors. The total net counts from MECS
(LECS) are $200 \pm 24$ ($88 \pm 14$) and represent about 36\% (47\%)
of the total (source + background) gross counts in the source region.


\begin{figure}
\psfig{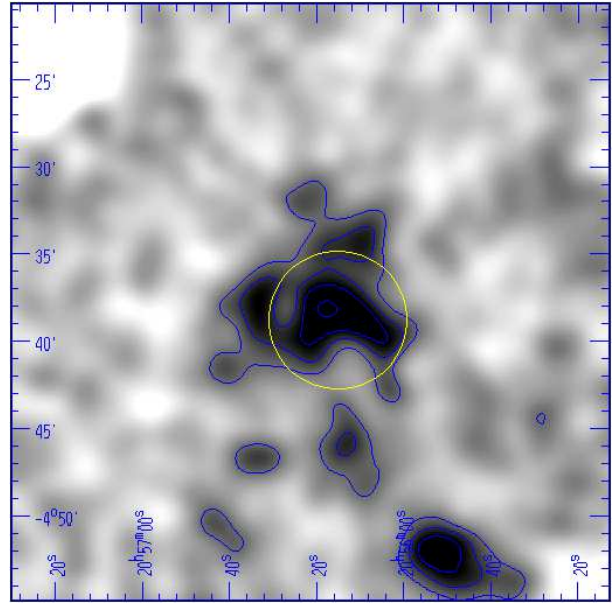}
\caption{MECS image in the $2-10$ keV band of the MS\thinspace
2053.7$-$0449 field ($35' \times 35'$). The raw data have been smoothed
with a two-dimensional gaussian filter with $\sigma = 0.8'$.  Contours
are at 3, 5, 7, 10 and 20 sigma above the background.  The white circle
represents the counts extraction region for MS\thinspace
2053.7.7$-$0449 (radius$= 4'$).}
\end{figure}

Table 3 reports the results of the spectral analysis.
The source spectrum was fitted using a Raymond-Smith 
spectral model modified by galactic absorption
($N_{H}=4.96\times 10^{20}$ cm$^{-2}$, Dickey \& Lockman 1990) along the
line of sight at the cluster position.  
From the present data no constraints can be set on the abundance which
was thus  fixed  to 0.3 solar.  The cluster gas is best
fit by a rest frame temperature $kT = 6.7^{+6.8}_{-2.3}$ 
(68 \% confidence interval).  
The unfolded spectrum, the folded spectrum  and the ratio between the
data and  best fit model are displayed in figure 6.

\begin{table*}
\begin{center}
\caption{Results of the Spectral Fit (LECS+MECS): Raymond-Smith Thermal Model.}
\begin{tabular}{lcccccc}
\hline
\hline
Name  & z & $N_{H_{Gal.}}$ & KT   & Abundance     & Norm. (LECS/MECS)$^{a}$ & $\chi^{2}_{\nu}$/dof  \\
& & ($10^{20}$ cm$^{-2}$) & (keV) & (Solar Units) & ($10^{-4}$) & \\
\hline
\hline
RXJ\thinspace 0152.7$-$1357  & 0.831  & 1.55 (fixed) &  6.46$^{+1.74}_{-1.19}$ & 0.53$^{+0.29}_{-0.24}$ 
                             & 6.9$^{+0.9}_{-0.8}$/9.7$^{+1.7}_{-1.5}$  & 0.79/21\\
MS\thinspace 2053.7$-$0449   & 0.583 & 4.96 (fixed) &  6.7$^{+6.8}_{-2.3}$ & 0.3 (fixed) 
                             & 5.4$^{+1.1}_{-0.9}$/4.8$^{+1.6}_{-1.1}$  & 0.93/7 \\
\hline
\hline
\end{tabular}
\end{center}
{NOTE - Errors are the 68\% confidence interval for one interesting parameter ($\Delta \chi^{2} = 1.0$).

$^a$ Normalization at 1 keV. This number is equal to 
$[10^{-14}/(4 \pi D^2)] \int n_e^2 dV$, where D is the distance to the 
source in cm, $n_e$ is the electron density in units of cm$^{-3}$
and V is the volume filled by the X-ray emitting gas in cm$^{3}$} 
\end{table*}

\begin{figure}
\hskip -1truecm
\psfig{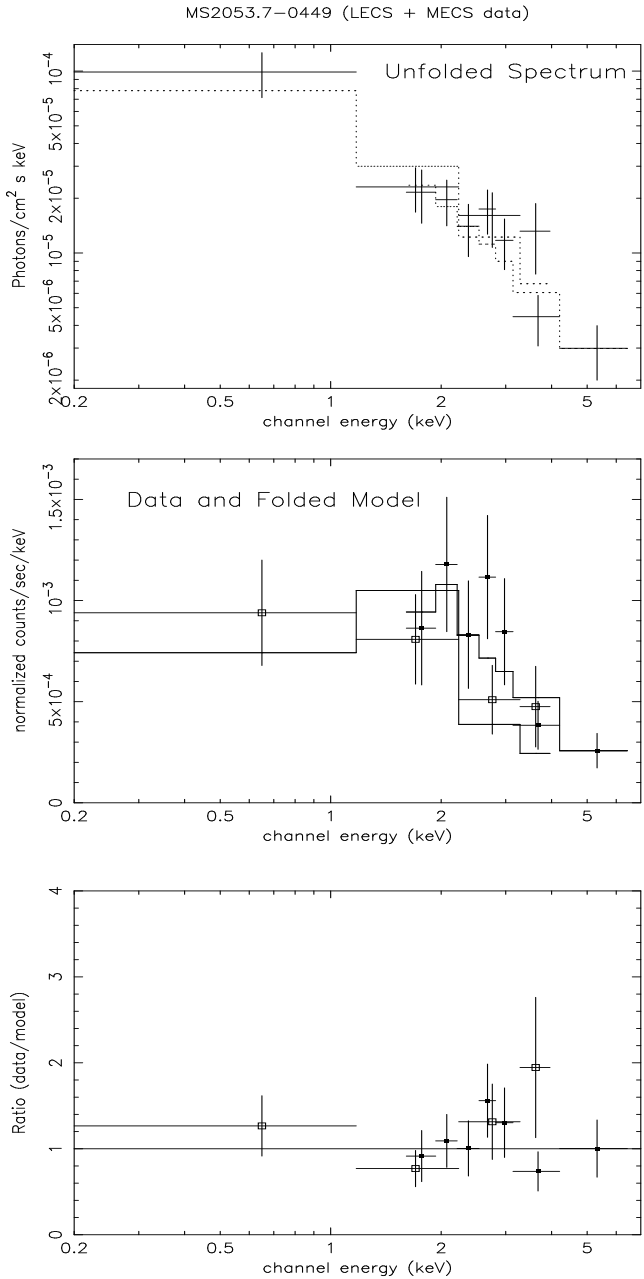}
\vskip 1.7truecm
\caption{MS\thinspace 2053$-$0449: Unfolded spectrum, folded spectrum and the 
ratio between data and best fit model.
In the last two panels we have indicated the LECS (MECS) data as open
(filled) squares}.
\end{figure}

The ratio of the model normalizations for LECS and MECS is about 1.1,
which is slightly in excess of the value expected from the known
differences in the absolute calibration of the two instruments.
However, given the present statistics we consider the fit 
acceptable.  We note that using only the MECS data we obtain for the
temperature $kT = 5.5^{+4.9}_{-1.75}$ (68\% confidence interval), which is
consistent with (and even more constrained than) that obtained by the
combined (LECS + MECS) data set.

Finally, we note that ASCA (GIS + SIS) 
did measure a temperature of $kT =
8.1^{+3.7}_{-2.2}$ (68\% confidence level) 
for MS\thinspace 2053.7$-$0449 (Henry, 1999); given
the large uncertainties involved (both from BeppoSAX and ASCA) the two
results are obviously consistent.  On the other hand they both suggest
a temperature less than about 13--14 keV.

Using the best fit model parameters reported in Table 3, 
we derive an unabsorbed ($0.3 - 3.5$) keV flux of 
$f_{0.3 - 3.5 keV} = (2.7 \pm 0.5) \times 10^{-13}$ erg cm$^{-2}$ s$^{-1}$ 
(LECS normalization) or $f_{0.3 - 3.5 keV} = (2.4 \pm 0.7) \times 10^{-13}$ erg 
cm$^{-2}$ s$^{-1}$ (MECS normalization).  The measured flux is 
lower, but consistent within the errors, 
than that derived from Einstein IPC data (see Table 1).
Henry (1999) finds from ASCA data a flux 
$f_{0.3 - 3.5 keV} = (3.85\pm 0.31) \times 10^{-13}$ erg cm$^{-2}$ s$^{-1}$ 
in perfect agreement with the EMSS.

The derived unabsorbed $2 - 10$ keV flux 
(computed assuming a mean value between the LECS and MECS normalizations)  
is 
$f_{2 - 10 keV} =  (1.9 \pm 0.5) \times 10^{-13}$ erg cm$^{-2}$ s$^{-1}$, 
corresponding to a luminosity in the cluster rest frame of 
$L_{2 - 10 keV} =  (3.9 \pm 1.0) \times 10^{44} \rm{h}_{50}^{-2}$ erg s$^{-1}$, 
and to a bolometric luminosity of 
$L_{bol} =  (8.2 \pm 2.2) \times 10^{44} \rm{h}_{50}^{-2}$ erg s$^{-1}$.

\section {Discussion}

In a recent work, Wu, Xue and Fang (1999) have used the largest sample
of clusters of galaxies with good broad band X-ray spectroscopy from
the literature to discuss the $L_{bol} - T$ and the $L_{bol} - \sigma$
relationships
\footnote{The reader is referred to Wu, Xue and Fang (1999) for the
comparison of their results with those from the literature, as well as for
the comparison of the observed relationships and the theoretical
ones.  See also Section 3.3 in Borgani et al. (1999).}.  
Their sample is mainly composed by clusters at
$z < 0.5$ (only 5 clusters out of 142 
are at higher redshift).  By comparing the clusters at $z < 0.1$ with
those at $z > 0.1$ they do not find convincing evidence for a
significant evolution in the $L_{bol} - T$ and the $L_{bol} - \sigma$
relationship out to $z\simeq 0.4$, a result 
which was first pointed out by Mushotzky and Scharf (1997).

As summarized in Table 4, distant cluster temperatures which have been
 measured to date are limited 
to a few systems at $z > 0.5$.  
At these redshifts the lookback time approaches half the age of the
Universe and, therefore, the time leverage to measure evolution in
cluster properties is large.


\begin{table*}
\begin{center}
\caption{Clusters at $z > 0.5$ with a Temperature Measurement.}
\begin{tabular}{llrrl}
\hline
\hline
Name                        & z       & KT$^a$                  & L$_{bol}$                          &  Ref.          \\
                            &         & (keV)                   & $10^{44} \rm{h}_{50}^{-2}$ erg s$^{-1}$   &                \\
\hline
\hline
MS\thinspace 0451.6$-$0305  & 0.539   & 10.17$^{+1.55}_{-1.26}$ & $53.7$                    & WXF,M+S,D96   \\
MS\thinspace 0015.9$+$1609  & 0.545   & 7.55$^{+1.0}_{-1.0}$    & $28.1$                    & WXF,M+S,HB98  \\
MS\thinspace 2053.7$-$0449  & 0.583   & 6.7$^{+13}_{-3.1}$      &  $8.2$                    & This paper    \\
MS\thinspace 1137.5$+$6625  & 0.78    & 5.7$^{+2.1}_{-1.1}$     & $16.0$                    & D99           \\
RXJ\thinspace 1716.6$+$6708 & 0.813   & 5.66$^{+2.54}_{-1.46}$  & $17.4$                    & WXF,G99       \\
MS\thinspace 1054.4$-$0321  & 0.826   & 12.30$^{+3.10}_{-2.20}$ & $19.9$                    & WXF,D98       \\
RXJ\thinspace 0152.7$-$1357 & 0.831   & 6.46$^{+3.2}_{-1.8}$    & $22.0$                    & This paper    \\
AXJ\thinspace 2019$-$1127   & 1.00    & 8.60$^{+9.5}_{-4.0}$    & $19.4$                    & WXF,H97       \\ 
\hline
\hline
\end{tabular}
\end{center}
{$^a$ For consistency with the other measurements we have reported in this column the 90\% confidence range.

Ref: WXF:  Wu, Xue and Fang (1999);
     M+S:  Mushotzky and Scharf (1997);
     D96:  Donahue (1996);
     D98:  Donahue et al. (1998);
     G99:  Gioia et al. (1999);
     D99:  Donahue et al. (1999);
     HB98: Hughes \& Birkinshaw (1998);
     H97:  Hattori et al. (1997).
      } 
\end{table*}

\begin{figure}
\psfig{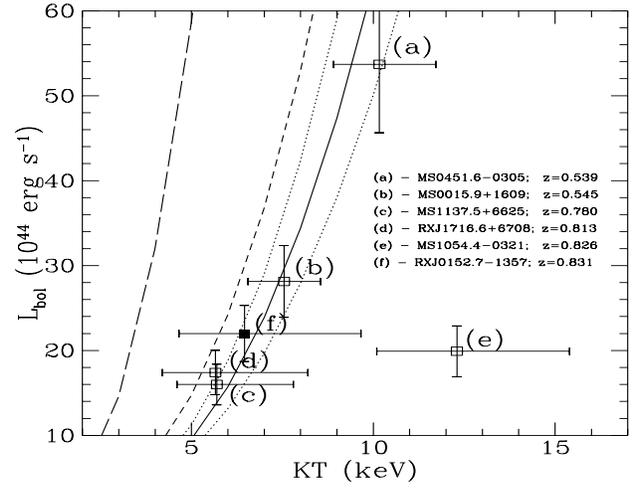}
\caption{Clusters at z $>$ 0.5 with a temperature measurement (see
Table 4).  The filled square represents the {\it Beppo}SAX measured
temperature of RXJ\thinspace 0152.7$-$1357 reported in this paper.  The
objects have been labeled in order of increasing redshift. 
Error bars on the temperature represent the 90\% confidence range,
while error bars on the bolometric luminosity represent a realistic
15\% absolute error. 
The solid line represents the
derived $L_{bol} - T$ relationship obtained by Wu, Xue and Fang (1999)
using a sample of 142 clusters from the literature; the dotted lines
represent the $\pm 2 \sigma$ scatter on the slope.  The short dashed
line represents the evolving $L_{bol} - T$ relationship with $(1+z)^1$
at $z=0.54$, while the long dashed line  represents the evolving
$L_{bol} - T$ relationship with $(1+z)^3$ at $z=0.83$. See section 5
for details.
}
\end{figure}

In figure 7, we plot the high redshift cluster temperatures known to
date in the $L_{bol}-T$ plane.  The error on the temperature represent
the 90\% confidence range, 
while we have used a realistic 15\% absolute
error for the bolometric luminosity.  Since MS\thinspace 2053.7$-$0449
and AXJ\thinspace 2019$-$1127
have very large errors on the measured temperature, we have not reported
these objects in figure 7.

The $L_{bol} - T$ relationship for
clusters at $z < 0.5$ obtained by Wu, Xue and Fang (1999) ($L_{bol} =
10^{-0.92\pm 0.05} T^{2.72\pm 0.05}$) is also shown (solid line),
together with the $\pm 2 \sigma$ on the slope
(dotted lines).
The short-dashed line represents the evolving $L_{bol} - T$
relationship with $L_{bol} \propto$ $(1+z)^{A1}$ at $z=0.54$, 
while the long dashed line represents the evolving
$L_{bol} - T$ relationship with $L_{bol} \propto$ $(1+z)^{A2}$ at
$z=0.83$.  The two redshifts enclose all the objects shown in figure
7. For A1 and A2 we have assumed the values 1 and 3, respectively; these two values
have been determined by Borgani et al. (1999), and  represent the 90\%
confidence level required to fit the lack of observed evolution of the
XLF in the RDCS cluster sample in a $\Omega_o$ = 1 universe.  Low
density models, instead, can easily be accommodated with a non-evolving
$L_{bol}-T$ relation, or mild ($A<1$) evolution.  The new cluster temperature
we have determined for RXJ0152.7$-$1357 is not consistent with a strong
evolution of the $L_{bol}-T$ relation out to $z\sim\! 0.8$.
Since all the data points in figure 7 lie to the right of the $A=1$ line,
according to the parameterization of Borgani et al. (1999), the cluster temperatures 
measured so far at $z>0.5$ lend considerable support to 
cosmological models with a low density parameter.
Similar results have been recently obtained by  Donahue et
al. (1999) and by  Donahue and Voit (1999).
Using a complete sample of high redshift EMSS clusters, Donahue et
al. (1999), have shown that the cluster temperature function reveals
modest evolution, a result which implies a low $\Omega$ value 
(Donahue and Voit 1999).

The metal abundance of the ICM in rich clusters of galaxies has been
recently investigated by Mushotzky and Loewenstein (1997).  They found
that the Fe abundance shows little or no evolution out to $z \sim 0.3$
($<Fe> \sim 0.3$), suggesting that most of the enrichment of the ICM
occurred at $z > 1$.  Given the present uncertainty on the ICM
abundance in RXJ\thinspace 0152.7$-$1357 we cannot set strong
constraints on the cosmological evolution of the Fe abundance.
However, within the large uncertainty ($A = 0.53^{+0.29}_{-0.24}$; 68
\% confidence interval), these data suggest that the bulk of the Fe
enrichment was completed by $z\sim 1$.

Finally it is worth comparing RXJ\thinspace 0152.7$-$1357 with
MS1054.4$-$0321. Provided that temperature measurements are not biased
by cooling flows, strong deviations from isothermality or by the
presence of contaminating AGN, it is interesting to note that the
temperature of MS1054.4$-$0321 is significantly higher than that of
RXJ\thinspace 0152.7$-$1357, although the two clusters have very
similar luminosities.  
With the present X-ray data we are unable to discuss any further the 
difference in temperature between these two distant clusters which have 
nonetheless many similarities. 
The highly spatially resolved X-ray spectroscopy and high throughput 
that {\it Chandra} and XMM will provide are needed to clarify this problem 
as well as to study in detail distant clusters of galaxies.


\section {Summary and Conclusions}

We have presented and discussed {\it Beppo}SAX  observations of two
high redshift clusters of galaxies: 
MS\thinspace 2053.7$-$0449 at $z = 0.58$
and RXJ\thinspace 0152.7$-$1357 at $z = 0.83$.

The first object has been selected from the EMSS sample.  No
constraints on the metallicity can be derived from the present data; by
fixing the abundance to 0.3 solar we measure a gas temperature with
very high uncertainty ($kT = 6.7^{+6.8}_{-2.3}$ keV).

The second cluster has been selected from the RDCS sample and is one of
the most X-ray luminous system known at $z>0.6$.  It is characterized
by a complex morphology (both in the optical and soft X-ray)  with at
least two cores, a gas temperature of $kT = 6.46^{+1.74}_{-1.19}$ keV
and a metallicity of $A = 0.53^{+0.29}_{-0.24}$ (68 \% confidence
interval). A prominent iron K$_{\alpha}$ line is clearly visible in the
MECS spectrum.

In light of this new cluster temperature measurement at $z \sim 0.83$, 
we have discussed the high redshift ($z > 0.5$) $L_{bol} - T$
relationship and we have found that the limited data available so far can
easily be accommodated with a non-evolving (or mildly evolving)
$L_{bol} - T$ relation.  This result, when combined with the little 
observed evolution in the X-ray cluster abundance, 
gives support to 
cosmological models with a low density parameter (e.g. Borgani et 
al., 1999).

\begin{acknowledgements}

RDC and IMG are greateful to T. Maccacaro for the continuous support and
encouragement since the early phases of this work. 
M. Hattori provided us with unpublished data for AXJ\thinspace 2019$-$1127. 
We thank G. Chincarini, C. Lobo, A. Wolter and G. Zamorani for a careful 
reading of the paper and  useful comments. 
We thank the {\it Beppo}SAX SDC, SOC and
OCC teams for the successful operation of the satellite and preliminary
data reduction and screening.  This research has made use of SAXDAS
linearized and cleaned event files produced at the {\it Beppo}SAX
Science Data Center.  IMG acknowledges partial financial support from
NSF AST95-00515, from NASA-STScI GO-06668.02-95A and from CNR-ASI
grants.

\end{acknowledgements}

\end{document}